# Single-shot wide-field optical section imaging

**Yuyao Hu,**[1,2] **Dong Liang,**[1,3] **Jing Wang,**[4,5] **Yaping Xuan,**[1,2] **Fu Zhao,**[1,2] **Jun Liu,** [1, 2*] **and Ruxin Li**[1,2]

[1] *State Key Laboratory of High Field Laser Physics and CAS Center for Excellence in Ultra-intense Laser Science, Shanghai Institute of Optics and Fine Mechanics, Chinese Academy of Sciences, Shanghai 201800, China*

[2] *University Center of Materials Science and Optoelectronics Engineering, University of Chinese Academy of Sciences, Beijing 100049, China*

[3] *School of Physics Science and Engineering, Tongji University, Shanghai 200093, China*

[4] *Institute of Photonic Chips, University of Shanghai for Science and Technology, Shanghai 200093, China*

[5] *Centre for Artificial-Intelligence Nanophotonics, School of Optical-Electrical and Computer Engineering, University of Shanghai for Science and Technology, Shanghai 200093, China*

*\*jliu@siom.ac.cn*

**Abstract:** Optical sectioning technology has been widely used in various fluorescence microscopes owing to its background removing capability. Here, a virtual HiLo based on edge detection (V-HiLo-ED) is proposed to achieve wide-field optical sectioning, which requires only single wide-field image. Compared with conventional optical sectioning technologies, its imaging speed can be increased by at least twice, meanwhile maintaining nice optical sectioning performance, low cost, and excellent artifact suppression capabilities. Furthermore, the new V-HiLo-ED can also be extended to other non-fluorescence imaging fields. This simple, cost-effective and easy-to-extend method will benefit many research and application fields that needs to remove out-of-focus blurred images.

## 1. Introduction

Wide-field imaging is a basic and popular imaging method owing to its high speed and low cost. But some common shortcomings are also obvious in wide-field imaging, in which the out-of-focus signal is usually blurred into background or noise, and the additional excitation of illumination light may reduce the signal-to-noise ratio(SNR) of the acquired image[1]. These shortcomings hinder the acquisition of images with high contrast[2, 3]. To reject this out-of-focus noise, a variety of optically sectioned techniques have been developed, such as confocal fluorescence microscopy[4], two-photon fluorescence microscopy[5], light-sheet fluorescence microscopy[6, 7], OS-SIM microscopy[8, 9], HiLo microscopy[10], and so on[11, 12]. Each technique has advantages and disadvantages owing to its special optical design. Among them, confocal or two-photon fluorescence microscopy based on laser scanning optical designs achieve optical-sectioned capability at the cost of low volume imaging speed. Light-sheet fluorescence microscopy needs expensive and inconvenient complex optical elements. Conventional OS-SIM microscopy can simplify the optical system, but three images with phase-shifted sinusoidal illumination patterns are often required to reconstruct an optically sectioned image, which restricts the imaging speed and the application of fast dynamic events[13].

Recently, the HiLo technique with excellent optical sectioning ability[1, 14-18] based on only two captured images, a uniform illuminated image and a structured illuminated image, has been widely used in conventional wide-field microscopy[19, 20], wide-field microendoscopy[21, 22], wide-field multiphoton microscopy[23], light-sheet microscopy[24-26], light field microscopy [27], multi-focus microscopy [28, 29] and incoherent holography [30], etc.

Even though the wide-field operating mode with only two captured images have already improved the imaging speed greatly, the needed two captured images in HiLo, a uniform illuminated image and a structured illuminated image, will still affect the improvement of both imaging speed and imaging quality. Firstly, it cannot make full use of the imaging speed potential of the scientific camera, because there is a switching between structured illumination and uniform illumination in HiLo. As a result, its imaging speed is always less than half that of the camera's imaging speed. Secondly, the requirement of two capture images means a compromise between imaging speed and imaging stability [31], otherwise artifacts will be introduced [16, 32]. As for fast moving objects, the moving induced artifacts between two captured images is unavoidable which will affect the final imaging quality. Thirdly, HiLo was rarely applied in label-free imaging, despite its outstanding sectioning capabilities. Inconvenient steps and redundant devices are part of the reason for the limitation. Therefore, there is no wide-field imaging method that can simultaneously meet all the requirements or advantages of fast, low cost, no artifacts, convenient and easy extension currently.

To overcome above constraints of existing optical sectioning techniques, inspired by the speckle contrast effect, a virtual HiLo (V-HiLo) based on edge detection (V-HiLo-ED) is proposed here to simultaneously simplify the optical system and improve both the imaging speed and quality. Only one uniform illumination wide-field image is required in this new method, which broke through the imaging speed limit of the conventional optical sectioning techniques such as HiLo or OS-SIM. With single image, there is no motion-induced artifacts. Furthermore, the optical sectioning performance is comparable to that of current optical sectioning methods. In this proof-of-principle research, by using several typical and irrelevant images, the results with proposed V-HiLo-ED is according with that using the conventional HiLo very well. Several experiments on either traditional wide-field label-free imaging or fluorescence imaging have also been executed which proved the effective of this new method.

Moreover, the proposed V-HiLo-ED exhibits many other advantages simultaneously, such as low-cost, simple setup, and low phototoxicity, etc., which make it a proud method for not only fluorescence microscopy but also other label-free imaging techniques such as incoherent digital holography, etc.

## 2. Principle of V-HiLo-ED

Different from conventional HiLo technique, which needs two images obtained by uniform illumination and structured illumination to reconstruct an optical- sectioning image, V-HiLo (virtual HiLo) only needs to capture a uniform illumination wide-field image to reconstruct the optically sectioned image, in which the structured illumination image is virtually obtained by a post-processing with the uniform illumination wide-field image.

According to the principle of conventional HiLo algorithm, the final optically sectioned image $I_{HiLo}(x,y)$ is reconstructed from two images together[33], the high frequency components $H_i(x,y)$ extracted from the uniform illumination image $I_u(x,y)$ and the low frequency components $L_o(x,y)$ extracted from the structured illumination image $I_s(x,y)$.

For clarity, the uniform illuminated image $I_u(x,y)$ can be decomposed into two parts, the in-focus component and the out-of-focus component, so $I_u(x,y)$ can be written as:

$$I_u(x,y) = I_{in}(x,y) + I_{out}(x,y) \tag{1}$$

Where $I_{in}(x,y)$ is the in-focus component, $I_{out}(x,y)$ is the out-of-focus component. The structured illumination image $I_s(x,y)$ is approximately described as [15]:

$$I_s(x,y) = \iiint PSF_{det}(x-x_o, y-y_o, z) Obj(x-x_o, y-y_o, z) S(x-x_o, y-y_o, z) dxdydz \tag{2}$$

Where $PSF_{det}$ is the point spread function of the detection arm, $Obj$ is object distribution. $S$ is the distribution of structured light on the object, and its contrast will decrease with defocus. The high-frequency in-focus component $H_i(x,y)$ can then be

expressed as:

$$H_i(x,y) = F^{-1}\{HP_{fc}\{F[I_u(x,y)]\}\} \quad (3)$$

Where $F$ and $F^{-1}$ are the Fourier and inverse Fourier transform operator. $HP_{fc}$ represents a 2D Gaussian high-pass filter in the spectral domain, where $f_c$ represents the cutoff frequency of the high-pass filter. For low-frequency in-focus component $L_o(x,y)$, it is can be expressed as:

$$L_o(x,y) = F^{-1}\{LP_{fc}\{F[W(x,y) \times I_u(x,y)]\}\} \quad (4)$$

Where $HP_{fc}$ represents a 2D Gaussian low-pass filter in the spectral domain where $f_c$ represents the cutoff frequency of the low-pass filter. $W(x,y)$ is a weighting map which is called contrast map. $W(x,y)$ can be shown as [20]:

$$W(x,y) = \frac{\sigma_\Lambda(x,y)}{\mu_\Lambda(x,y)} \quad (5)$$

Where $\Lambda$ is the length of the square sampling window, which depends on the cutoff frequency $k_c$ of the filter [33]. Here, $k_c$ can be defined as $k_c = 1/(2\Lambda)$ [20]. $\sigma_\Lambda(x,y)$ and $\mu_\Lambda(x,y)$ are the standard deviation and mean value of difference image $I_{dbp}(x,y)$ respectively. The equations are as follow:

$$\mu_\Lambda(x,y) = \frac{I_{dbp}(x,y) * N_\Lambda(x,y)}{N_k} \quad (6)$$

$$\sigma_\Lambda(x,y) = \sqrt{\frac{(I_{dbp}(x,y))^2 * N_\Lambda(x,y) - N_k \times (\mu_\Lambda(x,y))^2}{N_k - 1}} \quad (7)$$

Where $*$ represents correlation operation. $N_\Lambda(x,y)$ is a kernel used to calculate $W(x,y)$, $N_k$ is the sum of window matrix $N_\Lambda(x,y)$. The difference image $I_{dbp}(x,y)$ can described as:

$$I_{dbp}(x,y) = F^{-1}\{\{F[I_u(x,y) - I_s(x,y)]\} \times BFP(f_x, f_y)\} \quad (8)$$

Where $BFP(f_x, f_y)$ is a 2D Gaussian bandpass filter used to accelerate the noise decay so as to distinguish the in-focus and out-of-focus component. Its functional form is equivalent to subtraction of two low pass filters, which can be presented as [22]:

$$BFP(f_x, f_y) = \exp\left(-\frac{f_x^2 + f_y^2}{2\sigma^2}\right) - \exp\left(-\frac{f_x^2 + f_y^2}{\sigma^2}\right) \quad (9)$$

Where $(f_x, f_y)$ is coordinate at spectral domain, $\sigma$ is the standard deviation of the bandpass pass filter [20, 33]. After calculating based on the above parameters, the final HiLo image can then be reconstructed as:

$$I_{HiLo}(x, y) = H_i(x, y) + \eta L_o(x, y) \quad (10)$$

where $\eta$ is a custom scaling factor that can be adjusted experimentally to ensure the transition from low to high frequencies in $I_{HiLo}(x, y)$ occurs seamlessly [10, 24, 32]. The typical range of scaling factor $\eta$ in our experiment is $0 < \eta < 1$.

Here, edge detection is used in the post-processing in the V-HiLo-ED method, which is also a research hotspot in the field of computer vision. It is well known that the edge appears as a sharp local intensity change or discontinuity in the image [34]. Edge detection is not only suitable for segmentation of different objects [35], but also has satisfactory performance in image depth extraction [36]. So we can extract the edge of different targets in focus through edge detection.

The Laplacian of Gaussian (LOG) for edge detection was proposed by D. Marr and E. Hildreth in 1980 [37]. Since the Laplace operator detects the edge based on the zero-crossing point of the second derivative of the image, it is sensitive to discrete points and noise [38]. Considering that there are many discrete points and noises in the acquired wide-field image that are not conducive to the differential operation process in edge detection, the LOG is favored for its excellent anti-noise performance and rotation consistency. Therefore, LOG operation can be performed on the image and calculate its gradient to extract the image edge of the in-focus component [39], where the selection threshold is achieved through adaptive threshold algorithms such as Otsu threshold segmentation [40]. This parameter-free edge detection method can be extended to various targets [41]. As for reproducibility, LOG can be supported by Image Processing Toolbox of MATLAB.

The LOG can be expressed as $\nabla^2 G$, where $\nabla^2$ is the Laplacian operator (second order differential operator), and $G$ is a 2D Gaussian filter function. They can be expressed as:

$$\nabla^2 = \frac{\partial^2}{\partial x^2} + \frac{\partial^2}{\partial y^2} \tag{11}$$

$$G(x,y) = \frac{1}{\sqrt{2\pi\sigma^2}} \exp\left(-\frac{x^2+y^2}{2\sigma^2}\right) \tag{12}$$

Where $\sigma$ is standard deviation. In LOG method, $x$ and $y$ are coordinates at spatial domain. We need to perform a Gaussian filtering on the image $I(x, y)$ to reduce noise and discrete information before edge detection, which can be written as $I(x,y) \otimes G(x,y)$, where $\otimes$ is a convolution operation, and then we perform the filtered result with the Laplacian operator, the expression is as follow:

$$\nabla^2 [I(x,y) \otimes G(x,y)] \tag{13}$$

Furthermore, according to the property of derivative equation, the partial derivative of the Gaussian function can be performed firstly, and then convolute with the image $I(x,y)$. Then, the above expression can be described as:

$$\nabla^2 [I(x,y) \otimes G(x,y)] = \nabla^2 G(x,y) \otimes I(x,y) \tag{14}$$

For simplicity, the constant term of the Gaussian function can be omitted. The derivation and results of the second-order partial derivation of the Gaussian function are shown as follows:

$$\begin{aligned}
\nabla^2 G(x,y) &= \left(\frac{\partial^2}{\partial x^2} + \frac{\partial^2}{\partial y^2}\right) \times \exp\left(-\frac{x^2+y^2}{2\sigma^2}\right) \\
&= \frac{\partial}{\partial x}\left[\frac{-x}{\sigma^2} e^{\left(-\frac{x^2+y^2}{2\sigma^2}\right)}\right] + \frac{\partial}{\partial y}\left[\frac{-y}{\sigma^2} e^{\left(-\frac{x^2+y^2}{2\sigma^2}\right)}\right] \\
&= \left[\frac{x^2}{\sigma^4} - \frac{1}{\sigma^2}\right] e^{\left(-\frac{x^2+y^2}{2\sigma^2}\right)} + \left[\frac{y^2}{\sigma^4} - \frac{1}{\sigma^2}\right] e^{\left(-\frac{x^2+y^2}{2\sigma^2}\right)} \\
&= \left[\frac{x^2+y^2-2\sigma^2}{\sigma^4}\right] e^{\left(-\frac{x^2+y^2}{2\sigma^2}\right)}
\end{aligned} \tag{15}$$

Then, the imaging intensity based on the LOG edge detection can be expressed as:

$$I_L(x,y) = \nabla^2 G(x,y) \otimes I_u(x,y) \tag{16}$$

Where $I_L(x,y)$ contains the edge intensity of the objects in the focal plane, it has a similar effect to the conventional structured speckle illumination image in calculating local contrast, so $I_s(x,y)$ can be replaced by $I_L(x,y)$ to achieve virtual modulation, the formula is:

$$I_s(x,y) \approx I_L(x,y) = \nabla^2 G(x,y) \otimes I_u(x,y) \tag{17}$$

This is because the local speckle grain contrast of the conventional structured image is blurred or even disappears in the out-of-focus part, and the local speckle grain contrast in the focus part is clear [1]. Similarly, the edge detection image only has obvious edge information in the in-focus part (high local contrast), and retains weak intensity or even no signal in the out-of-focus part (low local contrast). In this way, V-HiLo-ED can extract a virtual structured illumination image from only a uniform illumination image based on edge detection, and then reconstruct a clear optically sectioned image.

The main process of V-HiLo-ED is shown in Fig. 1, and the corresponding sample images are attached to each step to show the change process. Note that the capability and reliability of the V-HiLo-ED was firstly proved by using some already published typical images using conventional optical sectioning method, as shown in the supplementation.

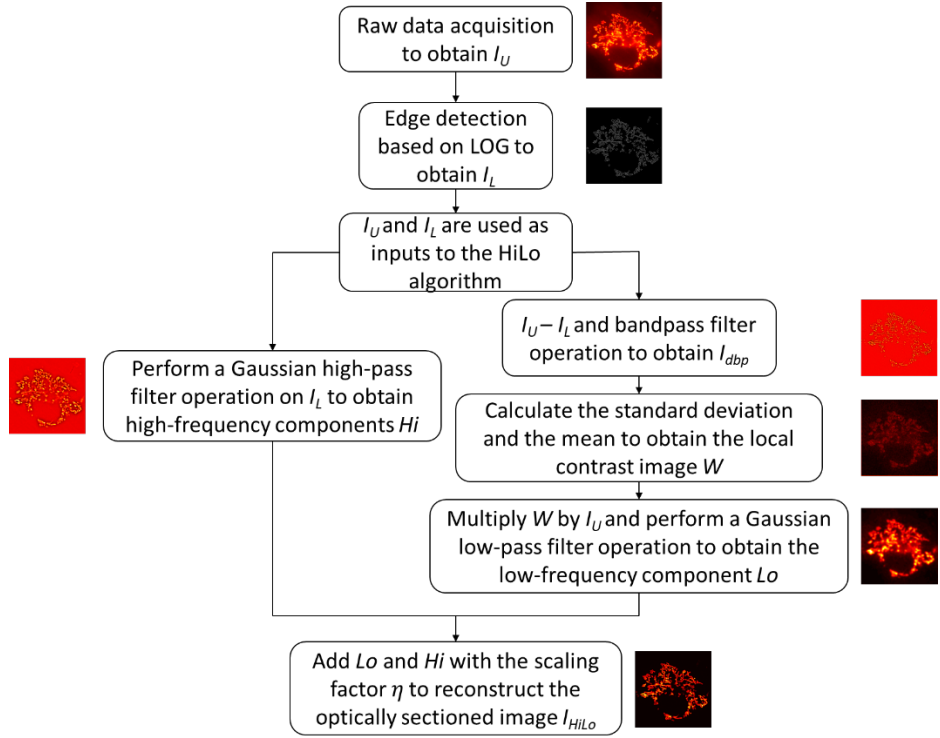

Fig. 1. Overview of the procedure for V-HiLo-ED.

## 3. Results and discussion

*3.1 Optical section fluorescent imaging with V-HiLo-ED*

Defocus background suppression greatly improves the SNR of biological images, which is of significance for observing organisms[13, 21, 42]. To characterize the optical sectioning capability of V-HiLo-ED, a single wide-field image of antibody-labeled mitochondria was used to reconstruct optically sectioned image through V-HiLo-ED for comparison (Fig. 2). In experimental setup, the wide-field fluorescent image was collected onto an EMCCD camera (Evolve 512, Delta Photometrics) with an oil-immersed objective (150×/1.45, Olympus), yielding a pixel size of 106 nm. Fig. 2(a) indicates a wide-field fluorescence image of the antibody-labeled mitochondria. After the LOG edge detection process, the in-focus edge image is shown in Fig. 2(b), which is used as the structured illumination image input of the HiLo algorithm. Fig. 2(c) is the Lo component determined together by the uniform illumination image (Fig. 2(a)) and the edge detection image (Fig. 2(b)). Fig. 2(d) is and the Hi component affected by the uniform illumination image (Fig. 2(a)) alone. In the experiment, they are added in an appropriate scaling factor to obtain the final optically sectioned image (Fig. 2(e)). The intensity distribution of the red dashed line at the same position of the wide-field image (Fig. 2(a)) and the optically sectioned image (Fig. 2(e)). From the results of the normalized intensity curve, it is obvious that there is almost no attenuation of the in-focus signal intensity, and the out-of-focus blur noise has been significantly suppressed. This means that the proposed method can achieve excellent optical section performance on antibody-labeled mitochondria that, to our knowledge, have not appeared in conventional HiLo microscopy previously.

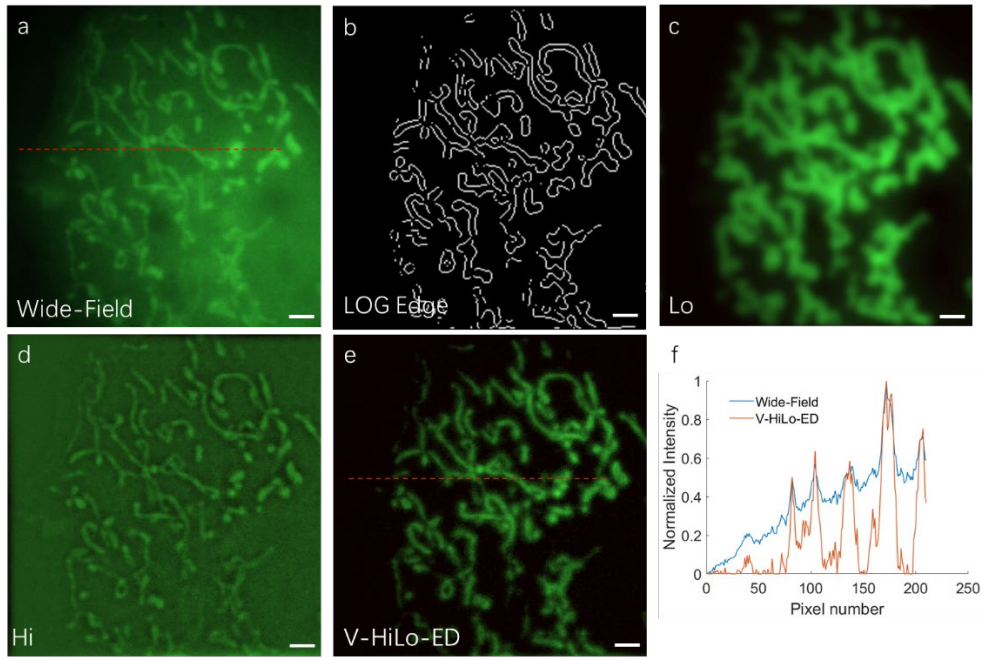

Fig. 2. Results of optical sectioning fluorescent imaging with V-HiLo-ED. (a) Uniform illuminated wide-field image of antibody-labeled mitochondria. (b) Edge detection image based on (a) and LOG operator. (c) Lo component image calculated based on (a), (b) and HiLo algorithm. (d) Hi component image calculated based on (a) and HiLo algorithm. (e) Optically sectioned image reconstructed based on (a), (b) and V-HiLo-ED method. (f) Normalized intensity profiles along red dash lines corresponding with (a) and (e). Scale bar: 2μm.

In subcellular imaging, high-resolution imaging(HR) is necessary, but due to the limitation of the field of view, a high dose light is often required to obtain a bright image [42], but this in turn burden more defocus noise [43], which may lead to worse image quality. Therefore, background suppression is particularly important in HR [44]. For this reason, two images with strong background noise, which are common in HR, are used to exemplify the background removal performance of V-HiLo-ED (Fig. 3). Fig. 3a and Fig. 3b are the wide-field image and the optically sectioned image reconstructed by V-HiLo-ED of COS-7 cell labeled with Tubulin-EGFP, respectively. More specifically, in Fig. 3a1 and Fig. 3b1, Fig. 3a2 and Fig. 3b2 are the magnified images of the yellow dashed box in Fig. 3a and Fig. 3b, respectively. The position pointed by the white arrows in the magnified images Fig. 3a1-a2 and Fig. 3b1-b2 show clearly that the separated microtubes can be clearly distinguished in the optically sectioned image after the V-HiLo-ED calculation. Fig. 3c and Fig. 3d are the antibody-labeled mitochondria images obtained by the wide-field image and the optically sectioned image reconstructed by V-HiLo-ED, respectively. Again, the out-of-focus noise has been well removed after the process of V-HiLo-ED. Similarly, Fig. 3c1-c2 and Fig. 3d1-d2 are the enlarged figures of the marked region with the yellow dashed box in Fig. 3c and Fig.7d, respectively. The small holes and gaps between the mitochondrial structures can also be clearly identified in the positions indicated by the white arrows in Fig. 3d1 and Fig. 3d2, which is almost invisible in the wide-field image shown in Fig. 3c or Fig. 3c1-c2. The experimental results once again demonstrate the excellent optical sectioning capability of V-HiLo-ED algorithm in HR. Some other typical technologies such as OS-SIM and traditional HiLo technology can also achieve this optical sectioning in HR with the cost of more image data acquisition, additional structured light lighting process[44],[45], iterative algorithms[46] and many incidental negative effects, which reflects that the proposed V-HiLo-ED method owns the advantages of higher flexibility, faster speed, simpler, more convenient and more economical benefits.

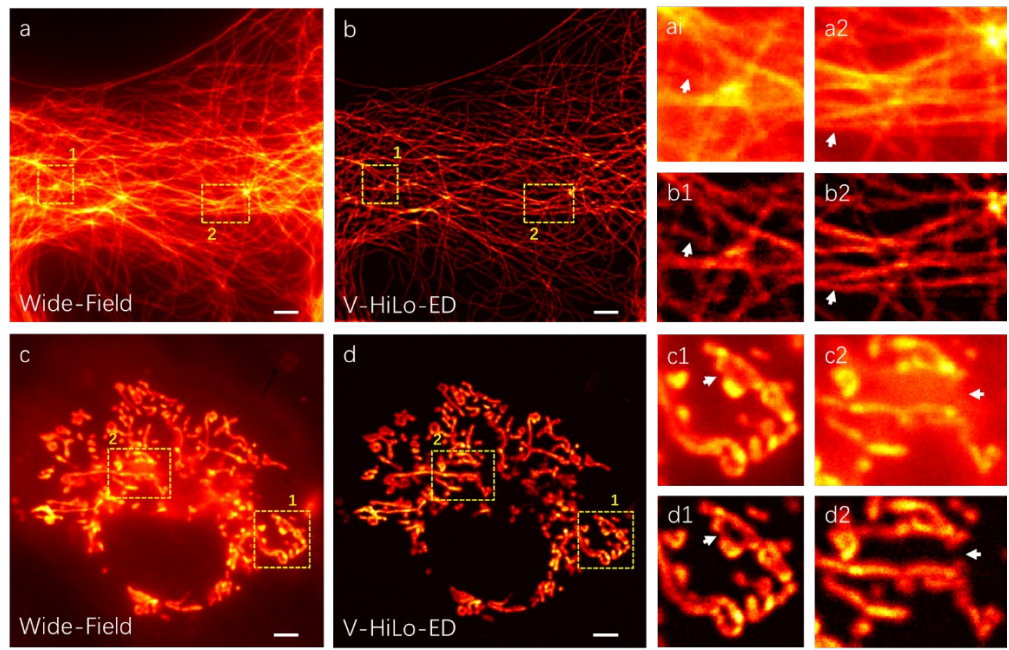

Fig. 3. Biological showcase of commonly of high-resolution imaging. (a) A COS-7 cell labeled with Tubulin-EGFP captured by typical wide-field microscopy. (b) Optically sectioned image of microtubule reconstructed by V-HiLo-ED. (c) antibody-labeled mitochondria captured by typical wide-field microscopy. (d) Optically sectioned image of mitochondria reconstructed by V-HiLo-ED. Scale bar: 2μm.

## 3.2 Optical section label-free imaging with V-HiLo-ED

A wide-field image of a standard resolution target (USAF 1951, Edmund) was used to achieve label-free optical section imaging and verify the resolution of V-HiLo-ED technology. The experimental setup is shown in Fig. 4. The light source from a LED is irradiated on two samples with different axial depths, and then converged on a CMOS camera for imaging. We can obtain a wide-field image with both defocus and in-focus components. Compared with the conventional optical section imaging experimental optical path, this setup is very cheap and extremely convenient, it hardly requires complicated adjustments, even collimation.

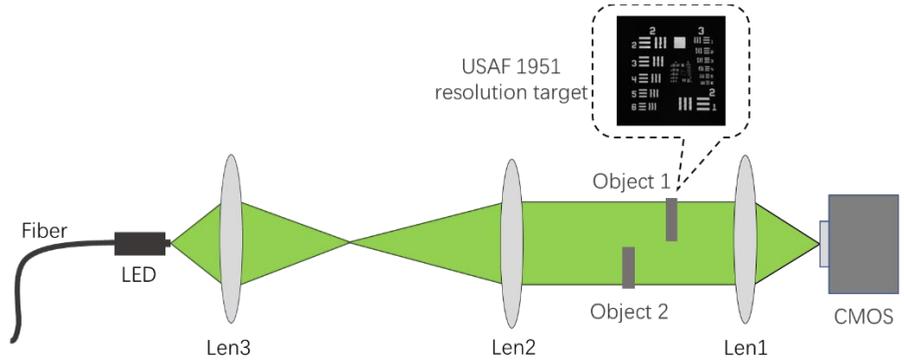

Fig. 4. Layout of optical sectioning label-free imaging with V-HiLo-ED. The LED with fiber provides incoherent light for uniform illumination. Len3 is a biconvex lens used to converge the beam. Len2 is a convex lens, which is used to make the beam irradiate the target 1 and target 2 at the same time, and the beam does not need to be collimated. Both object 1 and object 2 are standard resolution targets (USAF 1951, Edmond). Len1 is a convex lens used to converge the light beam onto the camera chip, the magnification of its imaging will vary with its focal length, and the focal length is selected based on the object being in the same field of view (FOV).

The experimental results are shown in Fig. 5. In the wide-field image (Fig. 5(a)), where object 1 is in focus and very clear, object 2 becomes blurry due to out of focus. After processing by the V-HiLo-ED algorithm, the reconstructed optical-sectioning image is shown in Fig. 5(b). The normalized one-dimensional intensity distributed along the horizontal red dashed lines in both the wide-field and the new V-HiLo-ED images are shown in Fig. 5(c). It shows clearly that the V-HiLo-ED image with better defocus blur suppression. To further confirm the spatial resolution of our method, the enlarged central part of the Object (USAF 1951 resolution target) with high resolution are shown in Fig. 5(d) and Fig. 5(e), respectively. Fig. 5(f) shows the normalized intensity distribution along the grating reticle (red dotted line). The results proved that the spatial resolution is almost the same, which has not been degraded. In summary, we provide an optical-sectioned image in label-free imaging without side effects. Which also means that V-HiLo-ED has a resolution similar to the raw wide-field image without compromising image quality.

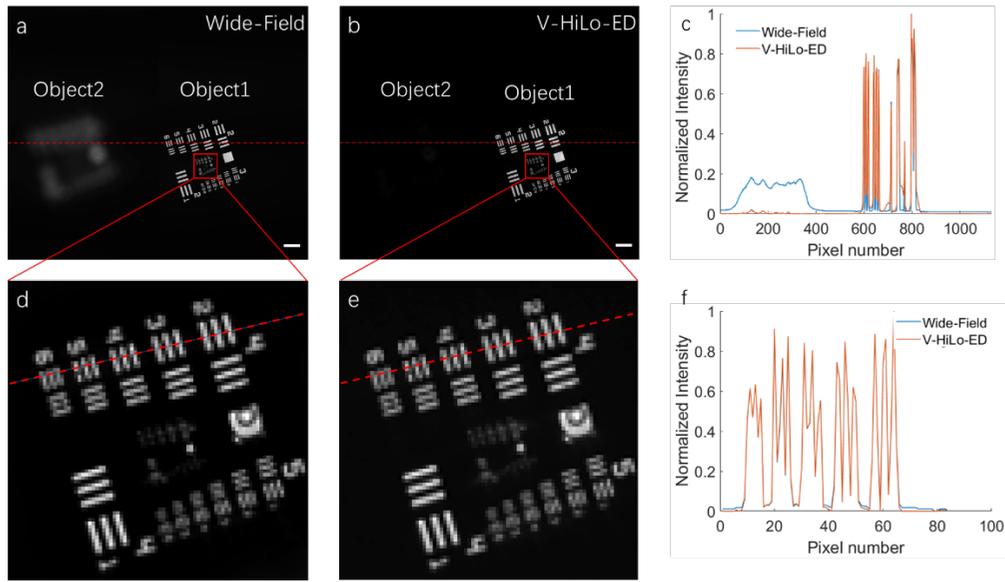

Fig. 5. Results of V-HiLo-ED in label-free imaging. (a) Uniform illuminated wide-field image of standard resolution targets. (b) Optically sectioned image reconstructed based on V-HiLo-ED. (c) Normalized intensity profiles along red dash lines corresponding with (a) and (b). (d) Enlarged central part of Fig(a). (e) Enlarged central part of Fig.(b). (f) Normalized intensity profiles along red dash lines corresponding with (d) and (e). Scale bar: 1mm.

### *3.3 V-HiLo-ED for Incoherent Digital Holography*

With its unique principles and advantages, holography is very popular in three-dimensional imaging and has brought some new applications [47]. However, due to various reasons, it has been troubled by defocusing and has not been solved very well. Such as incoherent digital holography [48], on the one hand, the introduction of structured illumination, whether modulated by grating patterns or spatial light modulators, will bring additional difficulties to holographic interferometric imaging that already requires strict adjustment accuracy, which reduces the feasibility of experiments. On the other hand, due to the low reconstruction clarity of incoherent digital holography [49], the uneven intensity and even the introduction of artifacts caused by structured illumination [50] will deteriorate the quality of the reconstructed result, which further increases the difficulty of achieving background suppression. Therefore, there is no application of optical sectioning technologies in incoherent digital holography until now. Fortunately, for V-HiLo-ED, the above restrictions can be cleverly avoided.

To verify the simplicity and applicability of our method, a pair of holographic reconstructed wide-field images in published paper[48] were used to calculate (Fig. 6). Fig. 6(a) and Fig. 6(b) are two images at different imaging depths. The upper right part of the image in Fig. 6(a) is in focus while the lower left part is out of focus, and the opposite for Fig. 6(e). Fig. 6(a) and Fig. 6(e) both use V-HiLo-ED for calculation, and the edge detection images are shown in Fig. 6(b) and Fig. 6(f) respectively, which show that the edge of the focal information is accurately extracted by the LOG operator. Fig. 6 (c) and Fig. 6(g) are the reconstructed optically sectioned images calculated by the V-HiLo-ED method, which are based on Fig. 6(a) and Fig. 6(e) respectively. By comparing the intensity distribution of the red dashed lines in Fig. 6(a) and Fig. 6(e), respectively, the results (shown in Fig. 6(d) and Fig. 6(h)) exhibit the optical-sectioning performance of V-HiLo-ED, which are not readily achieved by conventional optical sectioning methods in incoherent digital holography.

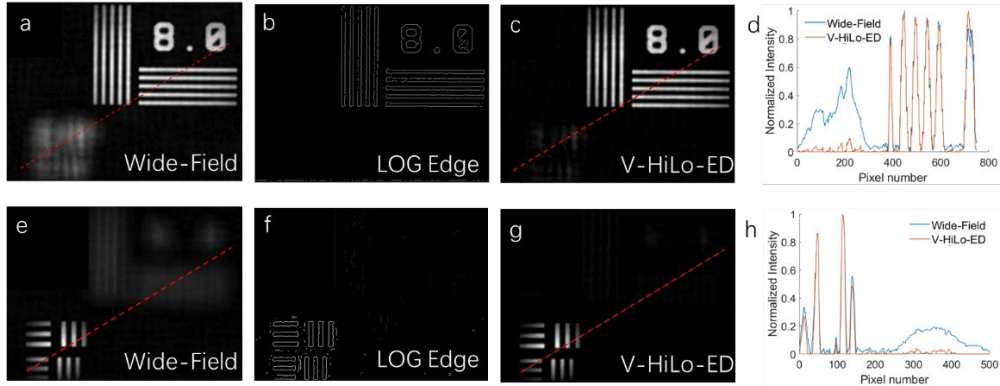

Fig. 6. Results of V-HiLo-ED for Incoherent Digital Holography (Ref. [48]). (a) (e) Two holographic reconstructed wide-field images at two different focal planes. (b) (f) Edge detection images based on LOG operator and (a) or (f), respectively. (c) (g) Optically sectioned image reconstructed based on V-HiLo-ED method, together with (a) and (b) or (e) and (f), respectively. (d)(h) Normalized intensity profiles along red dash lines corresponding with (a) and (c), or (e) and (g), respectively.

### 3.4 Suppressing motion-induced artifacts

In the conventional optical sectioning methods, such as OS-SIM[8] and HiLo[10], etc., at least two images are required. The imaging speed is generally less than half of the camera's capture speed. Furthermore, for fast-moving objects, two or more images are captured at different time which will cause motion-induced artifacts. Since the V-HiLo-ED technology requires simple single-shot image, the imaging speed is only limited by the camera. Besides increasing the imaging speed compared to conventional methods, the motion-induced artifacts can also be absolutely avoided. To show the advantage of V-HiLo-ED on motion artifact suppression more clear, existing data are used to simulate fast-moving objects for comparison.

Assuming that Fig. 7(a) and Fig. 7(b) are uniform illuminated images taken in two adjacent frames with delayed time, respectively. There is a translational in the longitudinal direction between Fig. 7(a) and Fig. 7(b) to simulate the fast movement of the object. Fig. 7(c) is the structured illumination image of Fig. 7(b). Based on the conventional HiLo algorithm, the Lo and Hi components calculated from Fig. 7(c) and Fig. 7(a) are shown in Fig. 7(d) and Fig. 7(e), respectively. Fig. 7(f) shows the final obtained optically sectioned image with obvious artifacts. This is because the movement of the object makes the captured two figures are different, which causes the calculated Lo component and Hi component to not be exactly paired [32]. And then the wrong optically sectioned image is reconstructed. As for V-HiLo-ED, these artifacts can be avoided because only a wide-field image is required. The calculated edge image from Fig. 7(a) is used as the structured illumination image (shown in Fig. 7(e)) of the HiLo algorithm. The calculated Lo component according to Fig. 7(g) is shown in Fig. 7(h), which in accord well with the Hi component that (shown in Fig. 7(e)) calculated from Fig. 7(a). With an appropriate scaling factor, the optically sectioned image based on V-HiLo-ED is shown in Fig. 7(i), which shows no artifacts (proved in supplementation).

As a result, the proposed new V-HiLo-ED method can effectively avoid the motion artifacts during imaging fast dynamic process that are unavoidable in the traditional optical sectioning methods due to the requirement of multiple frames of images.

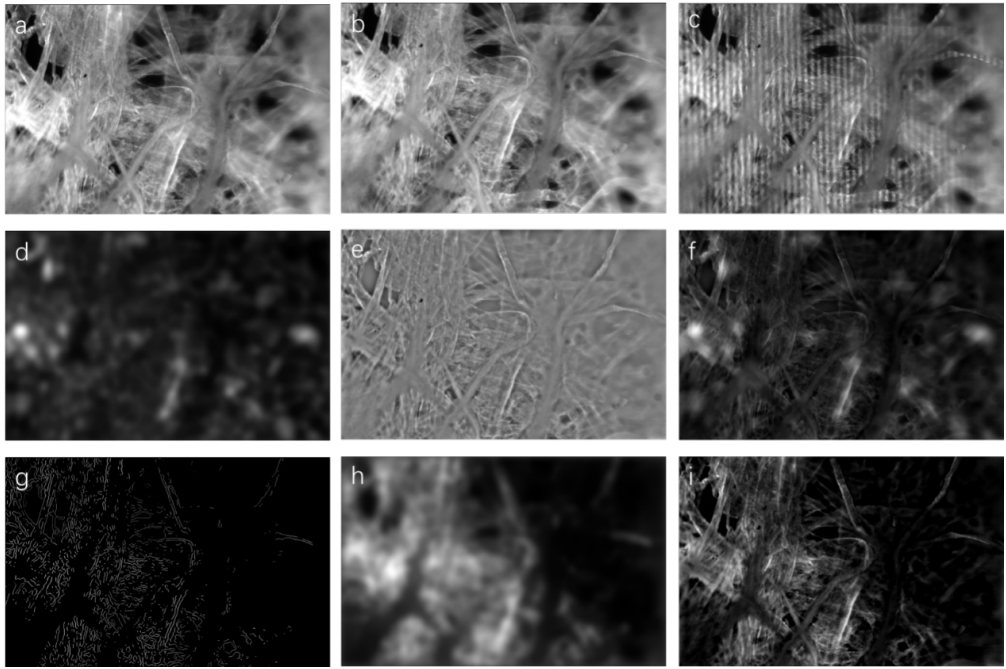

Fig. 7. V-HiLo-ED and conventional HiLo algorithms are used to perform optical section imaging of simulated fast-moving objects (Ref. [51]). (a) Uniform illuminated wide-field image at a certain location. (b) Uniform illuminated wide-field image with longitudinal translation relative to (a). (c) Structured illumination image based on (b). (d) Lo component image calculated based on (a), (c) and HiLo algorithm. (e) Hi component image calculated based on (a) and HiLo algorithm. (f) Optically sectioned image with obvious artifacts reconstructed based on (a), (c) and HiLo method. (g) Edge detection image based on (a) and LOG operator. (h) Lo component image calculated based on (a), (g) and HiLo algorithm. (i) Optically sectioned image reconstructed based on (a), (g) and V-HiLo-ED method.

## 4. Discussion and conclusion

Optical imaging has been widely applied in scientific researches, medical testing, and industrial gauging, where the imaging performance parameters such as imaging speed, field of view, imaging contrast, imaging quality, are all important that is hard to achieve at the same time. In the hot topic of fluorescence microscopy, the optical section imaging is very important to obtain a clear imaging with high contrast. So far, the previous proposed optical section imaging methods were almost all operated at the cost of imaging speed. Here, a novel V-HiLo-ED method is successfully applied to achieve wide-field optical section imaging with simply single uniform illumination, which can achieve high speed, wide-field, high fidelity imaging with simple setup and low cost at the same time. The capability and reliability were effectively verified by using several sets of typical and irrelevant imaging. The imaging capability and fidelity of the V-HiLo-ED method was well proved based on a label-free imaging, an incoherent digital holography imaging and a wide-field fluorescent imaging of antibody-labeled mitochondrial, which are rarely seen in conventional HiLo methods. Furthermore, the sample motion induced artifacts was simulated based on complex biological imaging, where artifacts are inevitable for conventional optical sectioning technology, but they can be ingeniously avoided for V-HiLo-ED.

In conclusion, we reported a single-shot wide-field optical section imaging technique based on edge detection combined with HiLo, which only requires a uniform illumination image to achieve excellent optical sectioning performance comparable to expensive optical sectioning microscopes such as confocal[1],[15],[31], etc. It not only improves the cost-effective, simplifies the complexity of the experiment, breaks through the speed limitation of the conventional optical sectioning methods, but also avoids various defects such as artifacts, phototoxicity, etc. carried by different optical sectioning technologies. Since the proposed V-HiLo-ED method can be easily extended to many unexpected fields, which is of great significance for promoting the application of this optical sectioning technology, we expected to further improve the V-HiLo-ED method and apply it to microendoscopy, multiphoton microscopy, and light-sheet microscopy, etc. in the near future.

## 5. Acknowledgments

The author would like to thank Prof. J. Mertz from Boston University for the helpful discussion on HiLo algorithm. This work was funded by the National Natural Science Foundation of China (NSFC) (61527821, 61521093, 61905257, U1930115), Shanghai Municipal Natural Science Foundation of China (20ZR1464500).

# 6. References

1. J. Mertz, "Optical sectioning microscopy with planar or structured illumination," Nat Methods **8**, 811-819 (2011).
2. C. Dussaux, V. Szabo, Y. Chastagnier, J. Fodor, J. F. Leger, L. Bourdieu, J. Perroy, and C. Ventalon, "Fast confocal fluorescence imaging in freely behaving mice," Sci Rep **8**, 16262 (2018).
3. J. Waters, "Sources of widefield fluorescence from the brain," Elife **9** (2020).
4. P. W. Hawkes, "Theory and Practice of Scanning Optical Microscopy - Wilson,T, Sheppard,C," Nature **312**, 384-385 (1984).
5. F. Helmchen, and W. Denk, "Deep tissue two-photon microscopy," Nat Methods **2**, 932-940 (2005).
6. J. Huisken, J. Swoger, F. Del Bene, J. Wittbrodt, and E. H. Stelzer, "Optical sectioning deep inside live embryos by selective plane illumination microscopy," Science **305**, 1007-1009 (2004).
7. B. C. Chen, W. R. Legant, K. Wang, L. Shao, D. E. Milkie, M. W. Davidson, C. Janetopoulos, X. S. Wu, J. A. Hammer, 3rd, Z. Liu, B. P. English, Y. Mimori-Kiyosue, D. P. Romero, A. T. Ritter, J. Lippincott-Schwartz, L. Fritz-Laylin, R. D. Mullins, D. M. Mitchell, J. N. Bembenek, A. C. Reymann, R. Bohme, S. W. Grill, J. T. Wang, G. Seydoux, U. S. Tulu, D. P. Kiehart, and E. Betzig, "Lattice light-sheet microscopy: imaging molecules to embryos at high spatiotemporal resolution," Science **346**, 1257998 (2014).
8. M. A. Neil, R. Juskaitis, and T. Wilson, "Method of obtaining optical sectioning by using structured light in a conventional microscope," Opt Lett **22**, 1905-1907 (1997).
9. J. Park, D. J. Brady, G. Zheng, L. Tian, and L. Gao, "Review of bio-optical imaging systems with a high space-bandwidth product," Advanced Photonics **3** (2021).
10. D. Lim, K. K. Chu, and J. Mertz, "Wide-field fluorescence sectioning with hybrid speckle and uniform-illumination microscopy," Opt Lett **33**, 1819-1821 (2008).
11. K. Patorski, M. Trusiak, and T. Tkaczyk, "Optically-sectioned two-shot structured illumination microscopy with Hilbert-Huang processing," Opt Express **22**, 9517-9527 (2014).
12. X. Zhang, Y. Chen, K. Ning, C. Zhou, Y. Han, H. Gong, and J. Yuan, "Deep learning optical-sectioning method," Opt Express **26**, 30762-30772 (2018).
13. Z. Li, Q. Zhang, S. W. Chou, Z. Newman, R. Turcotte, R. Natan, Q. Dai, E. Y. Isacoff, and N. Ji, "Fast widefield imaging of neuronal structure and function with optical sectioning in vivo," Sci Adv **6**, eaaz3870 (2020).
14. S. Santos, K. K. Chu, D. Lim, N. Bozinovic, T. N. Ford, C. Hourtoule, A. C. Bartoo, S. K. Singh, and J. Mertz, "Optically sectioned fluorescence endomicroscopy with hybrid-illumination imaging through a flexible fiber bundle," Journal of Biomedical Optics **14** (2009).
15. D. Lim, T. N. Ford, K. K. Chu, and J. Mertz, "Optically sectioned in vivo imaging with speckle illumination HiLo microscopy," J Biomed Opt **16**, 016014 (2011).
16. C.-Y. Lin, W.-H. Lin, J.-H. Chien, J.-C. Tsai, and Y. Luo, "In vivo volumetric fluorescence sectioning microscopy with mechanical-scan-free hybrid illumination imaging," Biomedical Optics Express **7** (2016).
17. H. Zhang, K. Vyas, and G. Z. Yang, "Line scanning, fiber bundle fluorescence HiLo endomicroscopy with confocal slit detection," J Biomed Opt **24**, 1-7 (2019).
18. W. Qiao, R. Jin, T. Luo, Y. Li, G. Fan, Q. Luo, and J. Yuan, "Single-scan HiLo with line-illumination strategy

for optical section imaging of thick tissues," Biomed Opt Express **12**, 2373-2383 (2021).

19. H. L. Fu, J. L. Mueller, M. P. Javid, J. K. Mito, D. G. Kirsch, N. Ramanujam, and J. Q. Brown, "Optimization of a widefield structured illumination microscope for non-destructive assessment and quantification of nuclear features in tumor margins of a primary mouse model of sarcoma," PLoS One **8**, e68868 (2013).

20. J. Schniete, A. Franssen, J. Dempster, T. J. Bushell, W. B. Amos, and G. McConnell, "Fast Optical Sectioning for Widefield Fluorescence Mesoscopy with the Mesolens based on HiLo Microscopy," Sci Rep **8**, 16259 (2018).

21. Q. Zhang, D. Pan, and N. Ji, "High-resolution in vivo optical-sectioning widefield microendoscopy," Optica **7** (2020).

22. H. Hsiao, C. Y. Lin, S. Vyas, K. Y. Huang, J. A. Yeh, and Y. Luo, "Telecentric design for digital-scanning-based HiLo optical sectioning endomicroscopy with an electrically tunable lens," J Biophotonics **14**, e202000335 (2021).

23. C. Y. Chang, C. H. Lin, C. Y. Lin, Y. D. Sie, Y. Y. Hu, S. F. Tsai, and S. J. Chen, "Temporal focusing-based widefield multiphoton microscopy with spatially modulated illumination for biotissue imaging," J Biophotonics **11** (2018).

24. J. Mertz, and J. Kim, "Scanning light-sheet microscopy in the whole mouse brain with HiLo background rejection," J Biomed Opt **15**, 016027 (2010).

25. D. Bhattacharya, V. R. Singh, C. Zhi, P. T. So, P. Matsudaira, and G. Barbastathis, "Three dimensional HiLo-based structured illumination for a digital scanned laser sheet microscopy (DSLM) in thick tissue imaging," Opt Express **20**, 27337-27347 (2012).

26. T. J. Schroter, S. B. Johnson, K. John, and P. A. Santi, "Scanning thin-sheet laser imaging microscopy (sTSLIM) with structured illumination and HiLo background rejection," Biomed Opt Express **3**, 170-177 (2012).

27. Z. Fu, Q. Geng, J. Chen, L. A. Chu, A. S. Chiang, and S. C. Chen, "Light field microscopy based on structured light illumination," Opt Lett **46**, 3424-3427 (2021).

28. C. Y. Lin, W. T. Lin, H. H. Chen, J. M. Wong, V. R. Singh, and Y. Luo, "Talbot multi-focal holographic fluorescence endoscopy for optically sectioned imaging," Opt Lett **41**, 344-347 (2016).

29. W. Lin, D. Wang, Y. Meng, and S. C. Chen, "Multi-focus microscope with HiLo algorithm for fast 3-D fluorescent imaging," PLoS One **14**, e0222729 (2019).

30. Y. H. Chia, J. A. Yeh, Y. Y. Huang, and Y. Luo, "Simultaneous multi-color optical sectioning fluorescence microscopy with wavelength-coded volume holographic gratings," Opt Express **28**, 37177-37187 (2020).

31. S. Kang, I. Ryu, D. Kim, and S. K. Kauh, "High-speed Three-dimensional Surface Profile Measurement with the HiLo Optical Imaging Technique," Current Optics and Photonics **2**, 568-575 (2018).

32. X. Zhou, P. Bedggood, and A. Metha, "Improving high resolution retinal image quality using speckle illumination HiLo imaging," Biomed Opt Express **5**, 2563-2579 (2014).

33. J. Mazzaferri, D. Kunik, J. M. Belisle, K. Singh, S. Lefrancois, and S. Costantino, "Analyzing speckle contrast for HiLo microscopy optimization," Opt Express **19**, 14508-14517 (2011).

34. Y. Y. Hung, Q. Zhu, D. Shi, and S. Tang, *Real-time edge extraction by active defocusing* (SPIE, 1991).

35. S. Lee, J. Y. Lee, W. Yang, and D. Y. Kim, "Autofocusing and edge detection schemes in cell volume measurements with quantitative phase microscopy," Opt Express **17**, 6476-6486 (2009).

36. K. Ko, and S. M. Abid Hasan, "Depth edge detection by image-based smoothing and morphological operations,"

Journal of Computational Design and Engineering **3**, 191-197 (2016).

38. D. Marr, and E. Hildreth, "Theory of edge detection," Proc R Soc Lond B Biol Sci **207**, 187-217 (1980).

38. A. K. Sao, and B. Yegnanarayana, "Edge extraction using zero-frequency resonator," Signal, Image and Video Processing **6**, 287-300 (2011).

39. X. Zhu, X. Li, K. Ong, W. Zhang, W. Li, L. Li, D. Young, Y. Su, B. Shang, L. Peng, W. Xiong, Y. Liu, W. Liao, J. Xu, F. Wang, Q. Liao, S. Li, M. Liao, Y. Li, L. Rao, J. Lin, J. Shi, Z. You, W. Zhong, X. Liang, H. Han, Y. Zhang, N. Tang, A. Hu, H. Gao, Z. Cheng, L. Liang, W. Yu, and Y. Ding, "Hybrid AI-assistive diagnostic model permits rapid TBS classification of cervical liquid-based thin-layer cell smears," Nature Communications **12** (2021).

40. V. M. Liarski, N. Kaverina, A. Chang, D. Brandt, D. Yanez, L. Talasnik, G. Carlesso, R. Herbst, T. O. Utset, C. Labno, Y. Peng, Y. Jiang, M. L. Giger, and M. R. Clark, "Cell distance mapping identifies functional T follicular helper cells in inflamed human renal tissue," Sci Transl Med **6**, 230ra246 (2014).

41. A. Safieddine, E. Coleno, S. Salloum, A. Imbert, A. M. Traboulsi, O. S. Kwon, F. Lionneton, V. Georget, M. C. Robert, T. Gostan, C. H. Lecellier, R. Chouaib, X. Pichon, H. Le Hir, K. Zibara, F. Mueller, T. Walter, M. Peter, and E. Bertrand, "A choreography of centrosomal mRNAs reveals a conserved localization mechanism involving active polysome transport," Nat Commun **12**, 1352 (2021).

42. X. Huang, J. Fan, L. Li, H. Liu, R. Wu, Y. Wu, L. Wei, H. Mao, A. Lal, P. Xi, L. Tang, Y. Zhang, Y. Liu, S. Tan, and L. Chen, "Fast, long-term, super-resolution imaging with Hessian structured illumination microscopy," Nat Biotechnol **36**, 451-459 (2018).

43. P. Dedecker, G. C. Mo, T. Dertinger, and J. Zhang, "Widely accessible method for superresolution fluorescence imaging of living systems," Proc Natl Acad Sci U S A **109**, 10909-10914 (2012).

44. Z. Weisong, Z. Shiqun, L. Liuju, H. Xiaoshuai, X. Shijia, Z. Yulin, Q. Guohua, H. Zhenqian, S. Yingxu, S. De-en, S. Chunyan, W. Runlong, Z. Shuwen, C. Riwang, X. Jian, M. Yanquan, W. Jianyong, J. Wi, C. Xing, D. Baoquan, L. Yanmei, M. Heng, S. Baoliang, T. Jiubin, L. Jian, L. Haoyu, and C. Liangyi, "Extending resolution of structured illumination microscopy with sparse deconvolution," Research Square (2021).

45. D. Dan, P. Gao, T. Zhao, S. Dang, J. Qian, M. Lei, J. Min, X. Yu, and B. Yao, "Super-resolution and optical sectioning integrated structured illumination microscopy," Journal of Physics D: Applied Physics **54** (2021).

46. S. Geissbuehler, A. Sharipov, A. Godinat, N. L. Bocchio, P. A. Sandoz, A. Huss, N. A. Jensen, S. Jakobs, J. Enderlein, F. Gisou van der Goot, E. A. Dubikovskaya, T. Lasser, and M. Leutenegger, "Live-cell multiplane three-dimensional super-resolution optical fluctuation imaging," Nat Commun **5**, 5830 (2014).

47. H. Hugonnet, Y. W. Kim, M. Lee, S. Shin, R. H. Hruban, S.-M. Hong, and Y. Park, "Multiscale label-free volumetric holographic histopathology of thick-tissue slides with subcellular resolution," Advanced Photonics **3** (2021).

48. J. Rosen, A. Vijayakumar, M. Kumar, M. R. Rai, R. Kelner, Y. Kashter, A. Bulbul, and S. Mukherjee, "Recent advances in self-interference incoherent digital holography," Advances in Optics and Photonics **11** (2019).

49. D. Liang, Q. Zhang, J. Wang, and J. Liu, "Single-shot Fresnel incoherent digital holography based on geometric phase lens," Journal of Modern Optics **67**, 92-98 (2020).

50. L. H. Schaefer, D. Schuster, and J. Schaffer, "Structured illumination microscopy: artefact analysis and reduction utilizing a parameter optimization approach," J Microsc **216**, 165-174 (2004).

51. D. L. a. J. Mertz, "HiLo imagej plugin," http://biomicroscopy.bu.edu/resources/.

**Supplement 1. Proving V-HiLo-ED with published images**

To test the capability, reliability and reproducibility of the proposed V-HiLo-ED method, several different images [1],[2] from already published papers based on either traditional HiLo or other optical section imaging methods have been executed by using the V-HiLo-ED method. The results are shown as follows.

The first published image is achieved by using the typical HiLo imaging technology applied to fluorescent biological samples, as shown in Fig. S1 (a)(b)(d). Fig. S1(a) is a representative complex image taken by the wide-field fluorescence microscopy provided by the J. Mertz team [1], it contains both the out-of-focus dense background and the in-focus imaging object. Fig. S1(b) is the structured illumination image with spatial modulated pattern. Many papers have demonstrated how to induce modulated pattern in the optical layout to achieve structured illumination[3-5]. In Fig. S1(b) with the structured illumination image, it is shown that local contrast is inversely proportional to the defocus distance. In some areas of the image with large defocus distance, the local contrast has even disappeared, while the focus area still maintains a high contrast. Through contrast-based calculations, the out-of-focus background can be suppressed, and the in-focus information is retained. Fig. S1(d) is the result obtained based on traditional HiLo, it is obvious that the background out-of-focus have been removed effectively.

As a comparison, V-HiLo-ED is performed based on single wide-field image of Fig. S1(a). Fig. S1(c) is the calculated image using the edge detection method based on the LOG operator and Fig. S1(a). According to the sharp change in the edge intensity of the focal plane, and the gently change in the defocus plane, the out-of-focus area in the image has been effectively suppressed after processing, and the edge of the in-focus area is well preserved. Fig. S1(e) shows image results based on the proposed V-HiLo-ED method. As we can see that the out of focus noise is eliminated, which is in according with Fig. S1(d) very well. To compare the results in Fig. S1(d) and Fig. S1(e) precisely, Fig. S1 (f) shows the one-dimensional intensity distribution of the red dashed line at the same location in Fig. S1 (d) and Fig. S1 (e). It can be seen that except for some small difference in intensity on some parts, the two lines show almost same precise structure.

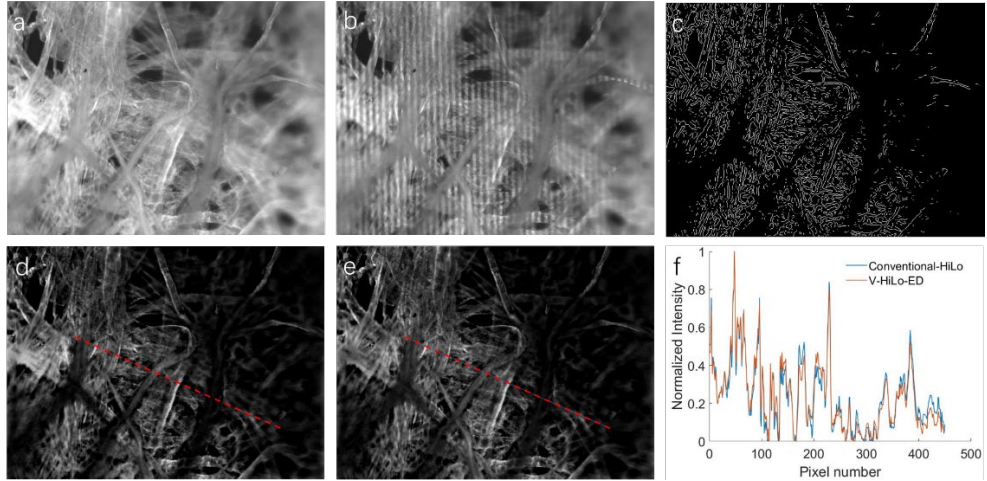

Fig. S1. Comparison of conventional HiLo and V-HiLo-ED with a typical sample image(Ref. [1]). (a) Uniform illuminated image. (b) Structured illumination image. (c) Calculated edge detection image based on (a) and LOG operator. (d) Optically sectioned image reconstructed based on (a), (b) and typical HiLo algorithm. (e) Optically sectioned image reconstructed based on (a), (c) and the proposed V-HiLo-ED method. (f) Normalized intensity profiles along red dash lines corresponding to (d) and (e).

In order to objectively and quantitatively assess the similarity of these two images Fig. S1(d) and (e), a common similarity metric: cosine similarity is chosen, which has been widely used in various fields [6-8] owing to its advantages [9] among many similarity metrics approaches. The cosine value is closer to 1, the more similar of the two images are. Here, the cosine values of the two HiLo images, they are Fig. S1(d) and Fig. S1(e), reaches 0.99941. The results prove that our V-HiLo-ED method based on single wide-field image shows a comparable performance to the conventional HiLo imaging method based on two captured images.

To avoid contingency, another publicly available data was selected for a further verification[2]. Fig. S2(a) shows the image of a sample, of which it changed from in-focus to out-of-focus from top toward the bottom. Fig. S2(b) and Fig. S2 (c) are obtained by using structured illumination or virtual detection based on edge detection, respectively. Accordingly, Fig. S2(d) and Fig. S2(e) are the calculated optically sectioned images based on the conventional HiLo method and the proposed V-HiLo-ED method, respectively. The two calculated images just like duplicate each other. Fig. S2 (f) shows the one-dimensional intensity distribution of the red dashed lines in both Fig. S2 (d) and Fig. S2 (e). The two lines overlapped each other very well, which highlights the ability of our V-HiLo-ED on suppressing out-of-focus signals is similar to that of the conventional HiLo technique.

Owing to simple single image is enough for the proposed V-HiLo-ED method, the virtual HiLo technique can improve the image speed of the traditional HiLo method. Furthermore, except for increase the image speed, as for fast moving objects, this single image detection method will absolutely remove the artifacts due to physical structure pattern [10] or object moving, which appear in many conventional methods such as SIM microscopy or HiLo method. As a result, the proposed method is very suitable for fast wide-field optical section imaging.

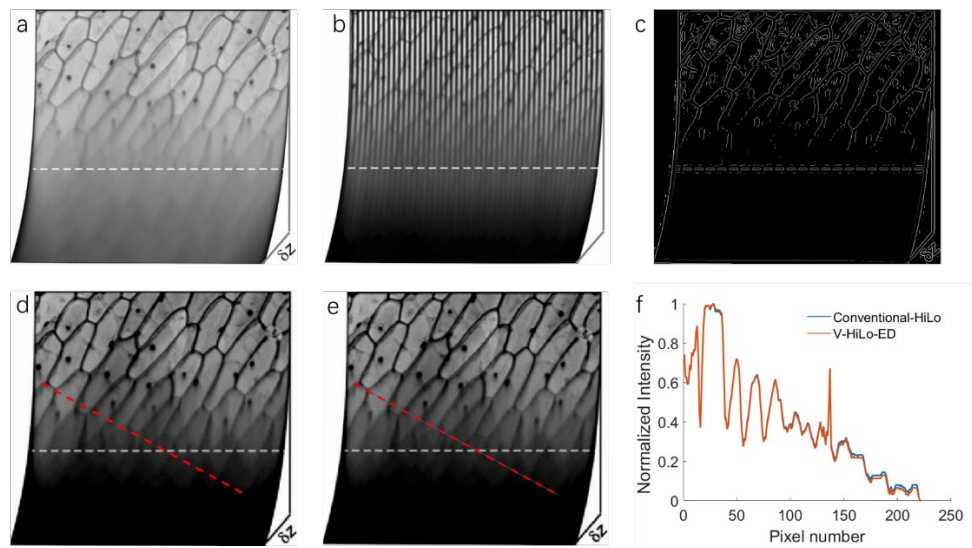

Fig. S2. Comparison of conventional HiLo and V-HiLo-ED with a onion biological image (Ref. [2]). (a) Uniform illuminated image. (b) Structured illumination image. (c) Calculated edge detection image based on (a) and LOG operator. (d) Optically sectioned image reconstructed based on (a), (b) and typical HiLo algorithm. (e) Optically sectioned image reconstructed based on (a), (c) and the proposed V-HiLo-ED method. (f) Normalized intensity profiles along red dash lines corresponding with (d) and (e).

### References of Supplement 1

1. D. L. a. J. Mertz, "HiLo imagej plugin," http://biomicroscopy.bu.edu/resources/.
2. D. Dan, B. Yao, and M. Lei, "Structured illumination microscopy for super-resolution and optical sectioning," Chinese Science Bulletin **59**, 1291-1307 (2014).
3. D. Bhattacharya, V. R. Singh, C. Zhi, P. T. So, P. Matsudaira, and G. Barbastathis, "Three dimensional HiLo-based structured illumination for a digital scanned laser sheet microscopy (DSLM) in thick tissue imaging," Opt Express **20**, 27337-27347 (2012).
4. S. Kang, I. Ryu, D. Kim, and S. K. Kauh, "High-speed Three-dimensional Surface Profile Measurement with the HiLo Optical Imaging Technique," Current Optics and Photonics **2**, 568-575 (2018).
5. H. Hsiao, C. Y. Lin, S. Vyas, K. Y. Huang, J. A. Yeh, and Y. Luo, "Telecentric design for digital-scanning-based HiLo optical sectioning endomicroscopy with an electrically tunable lens," J Biophotonics **14**, e202000335 (2021).
6. D. Wang, H. Lu, and C. Bo, "Visual Tracking via Weighted Local Cosine Similarity," IEEE Trans Cybern **45**,

1838-1850 (2015).

7. D. Sejal, T. Ganeshsingh, K. R. Venugopal, S. S. Iyengar, and L. M. Patnaik, "Image Recommendation Based on ANOVA Cosine Similarity," Procedia Computer Science **89**, 562-567 (2016).

8. H. Yu, S. Jia, Y. Liu, J. Peng, X. Zhou, and S. Yang, "Autofocusing based on cosine similarity in dual-wavelength digital holographic microscopy," Measurement Science and Technology **32** (2021).

9. P. Filev, L. Hadjiiski, B. Sahiner, H. P. Chan, and M. A. Helvie, "Comparison of similarity measures for the task of template matching of masses on serial mammograms," Med Phys **32**, 515-529 (2005).

10. L. H. Schaefer, D. Schuster, and J. Schaffer, "Structured illumination microscopy: artefact analysis and reduction utilizing a parameter optimization approach," J Microsc **216**, 165-174 (2004).